\documentclass[a4paper,12pt]{article}
\usepackage[utf8]{inputenc}
\usepackage[english]{babel}
\usepackage{amsmath, amssymb,graphicx}
\usepackage{caption}
\usepackage{subcaption}
\pdfoutput=1 
\usepackage{jheppubm} 
\usepackage{amsfonts}
\usepackage{ulem}

\usepackage{epsfig}

\newcommand{\nn}{\nonumber}
\newcommand{\be}{\begin{equation}}
\newcommand{\ee}{\end{equation}}
\newcommand{\bea}{\begin{eqnarray}}
\newcommand{\eea}{\end{eqnarray}}

\newcommand{\e}{\mathrm{e}}

\title{Holographic Dual to Conical Defects III: Improved Image Method}

\author[a]{I. Ya. Aref'eva,}
\author[a]{M. A. Khramtsov, 	}
\author[b]{M. D. Tikhanovskaya	}

\affiliation[a]{Steklov Mathematical Institute, Russian Academy of Sciences,\\Gubkina str. 8, 119991, Moscow, Russia}
\affiliation[b]{National Research Nuclear University ”MEPhI” (Moscow Engineering Physics Institute),
115409 Moscow, Russia}

\emailAdd{arefeva@mi.ras.ru}
\emailAdd{khramtsov@mi.ras.ru}
\emailAdd{tikhanovskaya@mi.ras.ru}

\abstract{ The geodesics prescription in holographic approach in Lorentzian signature is valid only for geodesics which connect spacelike-separated points at the boundary, since there are no timelike geodesics which reach the boundary. There is also no straightforward analytic Euclidean continuation for a general background, such as e. g. moving particle in AdS. We propose an improved geodesic image method for two-point Lorentzian correlators which is valid for arbitrary time intervals in case of the bulk spacetime deformed by point particles. We illustrate that our prescription is consistent with the case when the analytic continuation exists and with the quasigeodesics prescription used in previous work. We also discuss some other applications of the improved image method, such as holographic entanglement entropy and multiple particles in AdS$_3$. 
}
\begin{document}
\maketitle

\section{Introduction}
\label{sec:intro}
AdS/CFT correspondence and holographic approach \cite{Malda,GKP,Witten,Aharony1999} provide powerful methods for studying the dynamics of strongly-coupled systems in and out of equilibrium. Some of the systems admitting a holographic description are available for experimental study, such as quark-gluon plasma emerging from heavy-ion collisions \cite{Solana,IA,DeWolf}, as well as quantum fluids and superconductors \cite{Hartnoll08kx,Sachdev:2010ch}. 
This allows to use holographic techniques to gain insight into phenomena which lack systematic theoretical description, such as thermalization and quantum quenches \cite{Bal11,Lopez,Kera11,Caceres:2012em,Asplund2014,Bal12,Aref'eva:2013wma,IA15QGP}, transport properties of strongly-coupled systems \cite{Son:2002sd}, quantum entanglement problem \cite{Ryu:2006bv,AbajoArrastia:2010yt,Callan12,Bal14}, chaos and scrambling in QFT \cite{Maldacena15}. 
The holographic correspondence in its lower-dimensional form, namely AdS$_3$/CFT$_2$, allows to study $3D$ quantum gravity, in different regimes. Gravity in three dimensions is much simpler than in higher dimension, yet it still exhibits many of the key features of quantum gravity \cite{Maloney07}. Using the AdS$_3$/CFT$_2$ correspondence, one can probe it using powerful analytic techniques of two-dimensional conformal field theory \cite{Fitz14,Alkalaev15,Benjamin2016}. 

The geodesic approximation \cite{Bal99} plays a very important role in holographic calculations. It directly relates correlation functions of the boundary QFT to geometry of the bulk spacetime. In order to calculate the two-point correlator in the bulk between points $a$ and $b$ in semiclassical approximation, one has to sum over all bulk geodesics connecting these two points. Extrapolating these two points to the boundary via the BDHM prescription \cite{Banks1998}, one gets the two-point correlator in the boundary QFT. The geodesic prescription was used to describe the behaviour of physical quantities such as QCD Wilson loops during thermalization and quench, entanglement entropy and mutual information, see  \cite{Hubeny10,Bal11,Bal12,Lopez,Kera11,Caceres:2012em,Asplund2014,AbajoArrastia:2010yt,Callan12,Aref'eva:2013wma,IA15QGP,
Ryu:2006bv,Bal14,Albash:2010mv} and references therein. However, the geodesic prescription in its original form 
\cite{Bal99} is valid only either for Euclidean spacetimes, or for spacelike-separated points in the Lorentzian case. Timelike geodesics in asymptotically AdS spacetimes cannot reach the boundary, therefore the timelike region is unavailable to the prescription, unless there is an analytical continuation from the original Euclidean form. Because of this, the information carried by large-time dynamics, or even real-time correlators in general, cannot be obtained from the geodesics approximation in general spacetimes. The continuation of Lorentzian geodesic prescription has been considered before on certain locally AdS backgrounds. In \cite{Bal12} a non-trivial Euclidean continuation was constructed for the Vaidya spacetime. Another method that was used in \cite{Bal12,Arefeva2015} is making use of discontinuous timelike geodesics which go through Poincare horizon.

In the present paper we continue the work started in \cite{AAT,AA}. We propose the prescription for timelike correlations in locally AdS$_3$ spacetimes inspired by the latter method. We focus on the geodesic approximation of AdS$_3$ deformed by point particles. In the recent work \cite{AB-TMF,Arefeva2015,AAT,AA} the geodesic approximation for the boundary two-point function with spacelike-separated points was formulated in case of the AdS$_3$ spacetime with point particles. Particle solutions in AdS$_3$ \cite{Deser, Deser2,DeserLambda,Hooft} produce conical singularities, around which geodesics can wind. It was shown that the contribution of spacelike winding boundary-to-boundary geodesics to the correlator can be expressed as a sum over geodesics reaching to image points on the boundary\footnote{The geodesics approximation for the AdS-deficit spacetime has been compared with the holographic GKPW prescription in \cite{AK}.}. These images belong to orbit of the isometry transformation representing the topological identification (which in case of a particle in AdS$_3$ is the identification of faces of the wedge cut out by the particle). We propose to use a set of auxiliary geodesics with reversed causality relation between the endpoints to continue the geodesic correlators beyond the lightcone. This also allows to construct a continuation of images prescription formulated in \cite{AAT} to cases when the identification breaks causality at the boundary. We recover the pole structure of Lorentzian correlators by introducing appropriate factors by hand. The generalizations of this continuation of the image method are discussed; we show that it can be generalized to the case of AdS$_3$ deformed by multiple particles. Also we explain the application of the image method for calculation of the holographic entanglement entropy in the AdS with a moving conical defect.  

The paper is organized as follows. In section ~\ref{Setup} we introduce the ingredients of the image method, and reflection mapping on the boundary which is needed to continue our prescription into the timelike region. We proceed to formulate the prescription for two-point correlators in the entire boundary in section \ref{2point}. We illustrate the results of the prescription in the case of AdS$_3$ with a static and moving conical defects. In section \ref{Generals} we discuss some generalizations of our image method, in particular AdS$_3$ with multiple particles and a prescription for holographic entanglement entropy. The discussion of the results can be found in section \ref{conclusion}.

\section{Preliminary definitions \label{Setup}}

\subsection{The bulk spacetime}

We consider the AdS$_3$ spacetime with a particle inside. It can be obtained as a solution of Einstein equations in $3$ dimensions with a negative cosmological constant and a point particle stress-energy tensor. It does not perturb the metric locally and produces the conical singularity\footnote{Conical defects in the context of solids in flat $3D$ spacetime were considered also in \cite{volovich}. Holography with a conical defect in the boundary was studied e. g. in \cite{Bayona10}.} \cite{Deser,DeserLambda,Deser2,Hooft}, If the particle is static, the conical defect is located at the origin. Thus, static particle can be described by metric:
\bea
\label{barrA}\nn
ds^2&=&-\cosh^2\chi dt^2+d\chi^2+\sinh^2\chi d\varphi^2,\\
\varphi&\in& \left(0,2 \pi A \right).
\eea
Here $A=1-4G\mu \in (0,\ 1)$; $G$ is the Newtonian constant, $\mu$ is the mass of the particle ($\mu < 1/4G$ is assumed), $t$ is time coordinate and $\chi$ is radial coordinate (case of $\chi \rightarrow \infty$ corresponds to the conformal boundary). We will denote an angle of living space as $\bar{\alpha} : =2\pi A$ and the angle deficit by $\alpha : =2\pi(1-A)$.

Our main interest, however, lies in the spacetime with a moving particle in the bulk. Since the particle is massive, it cannot reach the boundary. One can show that the particle will move along a periodic worldline with period $T=2\pi$ \cite{Bal99}, which does not depend on the particle mass proportional to $\alpha$ or its rapidity $\xi$. It also generates a global defect in the spacetime, however the corresponding topological identification is more complex than simple angular identification. As shown in \cite{AAT}, one can obtain the identification isometry in the coordinate terms by boosting the wedge faces, which are cut out in the space by the particle. We denote the boost rapidity as $\xi$. The resulting transformation is
\bea
\tan t^*&=& \mathcal{B}_{\xi}(\alpha)\sec t  \tanh \chi \cos \varphi +\tan t \left(1+2\sinh^2\xi\sin^2\frac{\alpha}{2}\right)\,;\nn\\
\tan \varphi^* &=& - \frac{2\tan \varphi}{\mathcal{F}_\xi(\alpha)}\label{im3}\,; \eea
   \bea\label{chi}
\cosh\chi^*&=&\cosh\chi[ \left(\mathcal{B}_{\xi}(\alpha)\tanh \chi \cos \varphi +  \sin t (1+2\sinh^2\xi\sin^2\frac{\alpha}{2})\right)^2+ \cos ^2t ]^{\frac{1}{2}}.
    \eea
where $\mathcal{B}_{\xi}$ and $\mathcal{F}_\xi$ are defined as 
\bea\label{B-F}
\mathcal{B}_{\xi}(\alpha)&=&\sinh \xi\left( \sin \alpha  \tan \varphi
 -2\cosh\xi \sin^2\frac{\alpha}{2}\right),\\\nn
\mathcal{F}_\xi (\alpha)&=&\cosh \xi  (2 \sin \alpha \tan \varphi
   -\cos \alpha +\cos \varphi )\\&+&\sec \varphi \cos (\alpha +\varphi
   )+\cos \alpha \cosh2 \xi-2 \sinh ^2\xi\,;
\eea
To find the expression for the boundary isometry, induced by acting via $*$ in the bulk, we take the limit $\chi \to \infty$ in (\ref{im3}):
 \bea\label{nb-mov-iso}
 \tan t^* &=&\mathcal{B}_{\xi}(\alpha)\sec t \cos \varphi+\tan t \left(1+2\sinh^2\xi\sin^2\frac{\alpha}{2}\right)\,,\\
\tan \varphi^* &=& - \frac{2\tan \varphi}{\mathcal{F}_\xi(\alpha)} \nn\,.\eea
With the expression for the induced boundary identification at hand, one can now formulate the geodesic images prescription for spacelike geodesics, which was explained in \cite{Arefeva2015,AAT}. We will discuss it in the following subsection. 

\subsection{Winding geodesics and images\label{Geodesics}}

Having the identification isometry $*$ specified by (\ref{nb-mov-iso}), now we proceed to the description of the orbit of its boundary action. We also introduce the notation 
\be
\#=*^{-1} \label{diez} 
\ee
The coordinates of the boundary points obtained by acting with the isometry $n$ times are denoted as 
\bea
(\varphi,\ t)^{*\,n} &=&(\varphi^{*}_{n},\ t^{*}_{n})\,;\nn\\
(\varphi,\ t)^{\#\,n} &=&(\varphi^{\#}_{n},\ t^{\#}_{n})\,.\nn
\eea
Between the two given boundary points $a$ and $b$ one can have several geodesics. The geodesic that connects $a$ and $b$ directly, without crossing the wedge, is called basic. If it exists (recall that only spacelike-separated boundary points can be connected by a geodesic), its contribution to the correlator equals to 
 \be\label{G0}
\e^{-\Delta \mathcal{L}_{\text{ren}}(a, b)}=\left(\frac{1}{2(\cos (t_a-t_b)-\cos(\varphi_a-\varphi_b))}\right)^\Delta\,;
 \ee

This expression is periodic in $t \sim t+2\pi$,  This means that the contribution to the correlator from a geodesic between two spacelike points can be naturally continued along the time axis using this periodicity. To account for this continuation, we consider the lightcone which is now defined by the equation $\cos t = \cos \varphi$, rather than $t = \varphi$. From this point onwards, we consider points $x$ and $y$ of the boundary timelike-separated (or spacelike-separated), if we have $\cos (t_x-t_y) < \cos (\varphi_x - \varphi_y)$ (or $\cos (t_x-t_y) > \cos (\varphi_x - \varphi_y)$)
 
Other geodesics between $a$ and $b$ wind around the particle. It was shown in \cite{AB-TMF,AAT}, that their renormalized lengths can be expressed through the renormalized lengths of image geodesics\footnote{Provided the renormalization scheme respects identification isometry, see appendix and\cite{AAT}} connecting $a$ with $b^{*n}$ and $b^{\#n}$ (or, equivalently, $a^{*n}$ and $a^{\#n}$ with $b$). Thus, for spacelike-separated points the two-point correlator can be expressed as a sum over  direct geodesics between certain points in the AdS$_3$ spacetime with a piece removed by the wedge. 

 Of course, not all image points correspond to desired winding geodesics, so in general we have finite number of image geodesics contributing to the correlator. To distinguish the points in the sum over the isometry orbit which actually contribute, we define $\Theta$-functions. The definition is based on the geometric picture of AdS$_3$ as a cylinder. Then boundary points belong to the side surface of the cylinder, and in principle we can connect any two points at the boundary by drawing a straight line through the bulk of the cylinder. It is possible intersections of these lines with the wedge that define $\Theta$-functions.
The presence of contribution from the basic geodesic is regulated by the function $\Theta_{0}$. It is defined as:
\begin{itemize}
\item $\Theta_{0}(\varphi_x,t_x;\varphi_y,t_y;\alpha,\xi)=1$ if the straight line connecting points $(\varphi_x,t_x)$ and $(\varphi_y,t_y)$ does not cross the wedge;
\item $\Theta_{0}(\varphi_x,t_x;\varphi_y,t_y;\alpha,\xi)=0$ if the straight line connecting points $(\varphi_x,t_x)$ and $(\varphi_y,t_y)$ crosses the wedge.
\end{itemize}

To define functions which select image points for winding geodesics, we need some auxiliary notation. Denote by $w_+$, $w_-$ the faces of the wedge, such that under the isometries $*$ given by (\ref{im3}-\ref{chi}), and $\#$ given by (\ref{diez}) they transform as following:
\be
*:\, w_- \to w_+\,; \qquad \#:\, w_+ \to w_-\,.
\ee
Then functions $\Theta_{\pm}$ are defined as following:
\begin{itemize}
\item$ \Theta_{\pm}(\varphi_x,t_x;\varphi_y,t_y;\alpha,\xi)=1$ if the straight line from $(\varphi_x,t_x)$ to $(\varphi_y,t_y)$ crosses the face $w_{\pm}$ of the wedge first (or if one of the endpoints lie on $w_{\pm}$);
\item $\Theta_{\pm}(\varphi_x,t_x;\varphi_y,t_y;\alpha,\xi)=0$ if the straight line from $(\varphi_x,t_x)$ to $(\varphi_y,t_y)$ does not cross the wedge at all, or crosses $w_{\mp}$ first. 
\end{itemize}
An important remark is that our definition is somewhat different from that of \cite{AAT,AA}: to define the $\Theta$-functions we use straight lines instead of actual geodesics. This might seem a bit counter-intuitive, but actually the crossing of a geodesic between the given two points on the boundary of a particular wedge face is equivalent to the crossing of that face by the straight line connecting these two points. Unlike in \cite{AAT,AA}, now these definitions are not bound to the existence of actual geodesics between boundary points. Thus, we can use this definition for our generalization to the timelike separations right away.

\subsection{Reflection mapping and causality} 

If points $x$ and $y$ which belong to the boundary are timelike-separated, than there is no bulk geodesic between them. To generalize our geodesic prescription to this case, we use auxiliary spacelike geodesics.

\begin{figure}[ht]
\begin{center}
\includegraphics[width=7cm]{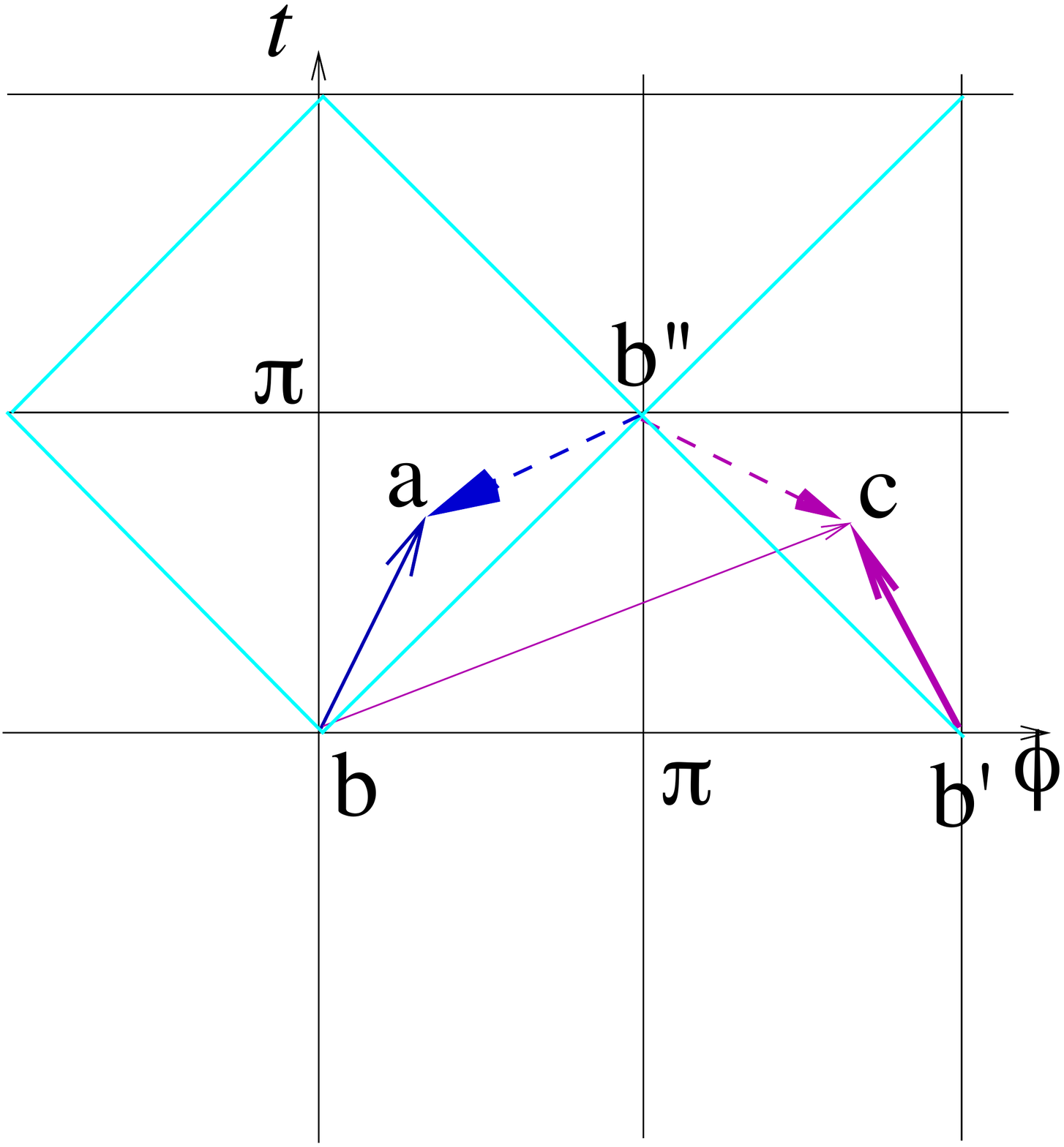}A\,\,\,\,\,
\includegraphics[width=7cm]{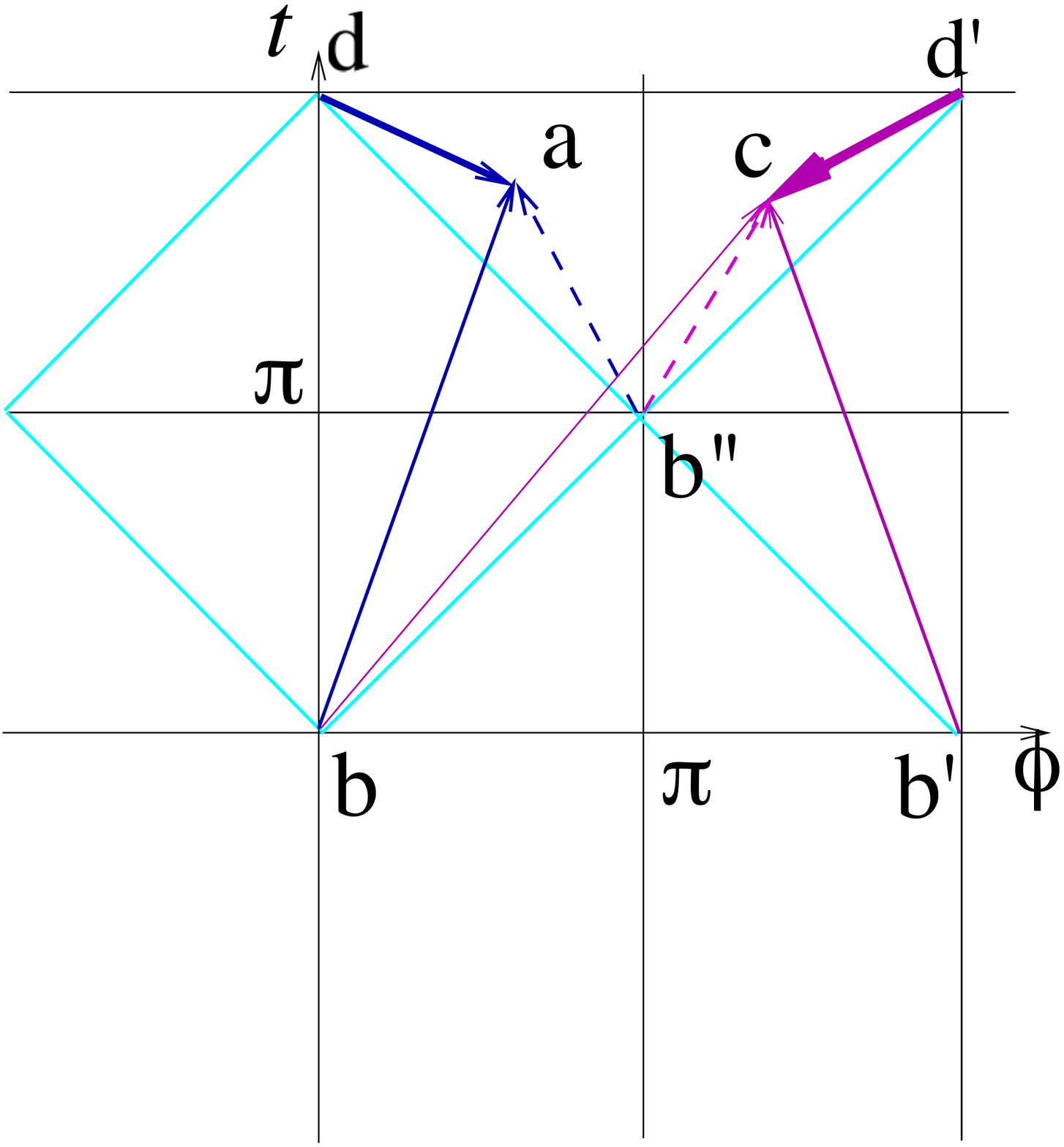}B
\caption{\textbf{A}. The plot of the reflection transformation of two vectors: $ \vec{ba}$ and $\vec{b^{\prime}c}$. After reflection transformation, accounting for the periodicity in angle: $ \vec{ba}\rightarrow \vec{b^{\prime \prime }a}$ (TL interval $\rightarrow$ SL interval), $ \vec{b^{\prime}c}\rightarrow \vec{b^{\prime \prime }c}$ (TL interval $\rightarrow$ SL interval) and $ \vec{bc}\rightarrow \vec{b^{\prime \prime }c}$ (TL interval $\rightarrow$ SL interval). \textbf{B}. The plot of the  reflection transformation  of two vectors: $ \vec{ba}$ and  $ \vec{b^{\prime}c}$. After reflection transformation, accounting for the periodic continuation along the time axis:  $ \vec{ba}\rightarrow \vec{b^{\prime \prime }a}$ or $ \vec{da}\rightarrow \vec{b^{\prime \prime }a}$ (SL interval $\rightarrow$ TL interval), $ \vec{b^{\prime}c}\rightarrow \vec{b^{\prime \prime }c}$ or  $ \vec{d^{\prime}c}\rightarrow \vec{b^{\prime \prime }c}$ (SL interval $\rightarrow$ TL interval) and $\vec{bc} \rightarrow \vec{b^{\prime \prime }c}$ or $ \vec{d^{\prime}c}\rightarrow \vec{b^{\prime \prime }c}$ (SL interval $\rightarrow$ TL interval).
}
\label{Fig:reflection}
\end{center}
\end{figure}

Here we introduce a useful notation to construct a prescription for the timelike correlations.
Define the map\footnote{The authors of \cite{HKLL06} used similar map, to which they referred as "antipodal map", to continue smearing functions from the Poincare patch into a region of global AdS.} acting at the boundary:
\be
R:\, a=(\varphi, t) \mapsto a^R=(\varphi+\pi, t+\pi)\,. \label{reflection}
\ee
This map reverses causal relation of the boundary points, provided we make the periodic continuation along the time direction: under this transformation the timelike interval $ab$, see Fig.\ref{Fig:reflection}, transforms to the spacelike one $ab''$ and vice versa. 
Instead of geodesic between timelike-separated points $x$ and $y$ we can now consider spacelike geodesic between points $x^R$ and $y$. Its contribution to the correlator equals to 
\bea
\e^{-\Delta \mathcal{L}_{\text{ren}}(x^R,\ y)} &=&Z \left(\frac{1}{2(\cos (t_x-t_y+\pi)-\cos(\varphi_x-\varphi_y+\pi))}\right)^\Delta\nn\\ &=& Z\left(\frac{1}{2(-\cos (t_x-t_y)+\cos(\varphi_x-\varphi_y))}\right)^\Delta\,;
\eea
where Z is the numerical factor emerging from the renormalization scheme (see appendix \ref{renormalization}).
Note that the expression in denominator in the parenthesis is positive, if $x$ and $y$ are timelike-separated. 

Since we plan to use our prescription to obtain the expression for two-point correlators in the coordinate representation for the entire Lorentzian boundary plane, we have to specify the causal structure of Lorentzian correlation functions. We first consider the Wightman correlator of scalar boundary operators $\langle \mathcal{O}_\Delta(\varphi_a,t_a)\mathcal{O}_\Delta(\varphi_b,t_b)\rangle$. We note that the Wightman CFT correlator on a cylinder can be written as \cite{OS,Luscher}:
 \bea
 \langle \mathcal{O}(t, \varphi)\mathcal{O}(0,0)\rangle &=& \left(\frac{1}{2\,\left(\cos (t-i\epsilon)  - \cos \varphi\right)}\right)^\Delta\label{Gw_unfolded} \\&=&
\left(\frac{1}{2\,\left|\cos t  - \cos \varphi\right|}\right)^\Delta\,e^{-i\,\pi\,\Delta \,\cdot\,\theta(-\cos t  + \cos \varphi)\,{\mbox {sign}}(\sin t)} \nn\\
&=&\left(\frac{1}{2\,\left|\cos t  - \cos \varphi\right |}\right)^\Delta\,
\left\{
\begin{array}{ccc}
 e^{-i\,\pi\,\Delta \,\cdot\,{\mbox {sign}}(\sin t _1)} & \,{\mbox {for timelike} } \\
 \, & \,  & \  \\
  1&  \,{\mbox {for spacelike} }
\end{array}
\right.\nn
\eea
From the Wightman correlators, using the standard QFT definitions, one can obtain the causal and retarded Green's functions. The retarded function is 
\bea
G_{ret}(t, \phi)&=&-\frac{2i\sin[ \pi \Delta \,{\mbox{sign}}(\sin t_1)] }{|\cos t_1  - \cos \phi_1|^\Delta}\,\theta(t_1)\theta (-\cos t_1  + \cos \phi_1)\\
&=&\left(\frac{1}{2\,\left|\cos t  - \cos \varphi\right |}\right)^\Delta\,\label{Gretarded}
\left\{
\begin{array}{ccc}
 -2^{\Delta-1} i \sin[\pi \Delta \text{sgn}(\sin t)] \theta(t)& \,{\mbox {for timelike} } \\
 \, & \,  & \  \\
  0&  \,{\mbox {for spacelike} }
\end{array}
\right.\nn
\eea
The Feynman propagator reads 
 \bea
 &&\langle T\mathcal{O}(t, \varphi)\mathcal{O}(0,0)\rangle = \left(\frac{1}{2\,\left(\cos (t-it\epsilon)  - \cos \varphi\right)}\right)^\Delta\label{Gc_unfolded} =\\
&&=\left(\frac{1}{2\,\left|\cos t  - \cos \varphi\right |}\right)^\Delta\,
\left\{
\begin{array}{ccc}
e^{-i\,\pi\,\Delta \,\text{sgn}(\sin t _1)}+\theta(-t)e^{+i\pi\,\Delta \,\text{sgn}(\sin t _1)} & \,{\mbox {for timelike} } \\
 \, & \,  & \  \\
1&  \,{\mbox {for spacelike} }
\end{array}
\right.\nn
\eea
The prefactor in front of the curly bracket of every correlator equals to $\exp(-\Delta \mathcal{L}_{\text{ren}}(t,\ \varphi;\ 0,\ 0))$ for spacelike case and $\exp(-\Delta \mathcal{L}_{\text{ren}}(t+\pi,\ \varphi+\pi;\ 0,\ 0))$ for timelike case. Thus, reflection geodesic prescription catches the behaviour of Lorentzian correlators without taking into account the causal structure of the correlation function. In the final image method formula, we therefore need to reintroduce the quantities after the bracket by hand. 

We also note that there is an alternative to the reflection geodesics - quasigeodesics \cite{Bal12,AAT}, which are discontinuous curves consisting of pieces of spacelike geodesics and a null geodesic. They give the same result and also do not carry any information about the causality of the two-point function. 

\section{Prescription for two-point correlators\label{2point}}

\subsection{General formula}

We now have all the tools to construct the Lorentzian two-point correlators of scalar operators with conformal dimension $\Delta$ between points $a$ and $b$ with arbitrary spacetime separation. Let the index $A$ denote the Wightman (W), causal (c) or retarded (ret) correlator. We propose the general form:
\bea\label{g-main}
 G_\Delta^A (t_a, \varphi_a; t_b, \varphi_b) &=& G_{\Delta,0} (\varphi_a,t_a;\varphi_b ,t_b)\,\Theta_{0}(\varphi_a,t_a;\varphi_b,t_b)\\&+&\sum_{n}\,G^A_{\Delta,n}(\varphi_a,t_a;\varphi^*_{b,n},t^*_{b,n})\,\Theta_{+}(\varphi_a,t_a;\varphi^*_{b,n},t^*_{b,n})\nn\\
&+& \sum_{n}\,G^A_{\Delta,n}
( \varphi_a,t_a;\varphi^{\#}_{b,n},t^{\#}_{b,n})\,\Theta_{-}(\varphi_a,t_a;\varphi^{\#}_{b,n},t^{\#}_{b,n}),\nn
\eea
where functions $G^A_{\Delta, n}$ are expressed through $G^A_{\Delta,0}$ times renormalization factors (\ref{G-ren-definition}), and $G^A_{\Delta,0}$ is defined as following:
\begin{itemize}
\item If points $x$ and $y$ are spacelike separated points
\bea
&& G^W_{\Delta,0}(\varphi_x,t_x; \varphi_y,t_y)=G^c_{\Delta,0}(\varphi_x,t_x; \varphi_y,t_y) = \e^{-\Delta \mathcal{L}_{\text{ren}}(x, y)}\,,\nn\\
&& G^{ret}_{\Delta,0}(\varphi_x,t_x; \varphi_y,t_y)=0\,;
\eea
where $\mathcal{L}_{\text{ren}}(x, y)$ is the renormalized length of the geodesic between points $x$ and $y$. The renormalization scheme is described in the appendix \ref{renormalization}.
\item If $x$ and $y$ are timelike separated points 
\bea
&& G^W_{\Delta,0} (\varphi_x,t_x; \varphi_y, t_y) = \e^{-\Delta \mathcal{L}_{\text{ren}}(x^R, y)}\e^{-i \pi\Delta\ \text{sgn} (t_x - t_y)}\,,\\
&& G^c_{\Delta,0} (\varphi_x,t_x; \varphi_y, t_y) = \e^{-\Delta \mathcal{L}_{\text{ren}}(x^R, y)}\left(\theta(t_x-t_y) e^{-i\,\pi\,\Delta \,\text{sgn}(\sin( t_x-t_y))}\right.\nn\\&&\left.\qquad\qquad\qquad\qquad+\theta(t_y-t_x)e^{+i\pi\,\Delta \,\text{sgn}(\sin (t_x-t_y))}\right)\,,\nn\\
&& G^{ret}_{\Delta,0}(\varphi_x,t_x; \varphi_y,t_y)= -\e^{-\Delta \mathcal{L}_{\text{ren}}(x^R, y)} \times 2^{\Delta-1} i \sin[\pi \Delta \text{sgn}(\sin (t_x-t_y))] \theta(t_x -t_y)\,;\nn
\eea
where $(\varphi_{x^R}, t_{x^R}) = (\varphi_{x}+\pi, t_{x}+\pi)$ - the image of the point $a$ under the reflection mapping $R$, which shifts the boundary coordinates by $\pi$. 
\end{itemize}
The $\Theta$-functions defined in \ref{Geodesics} provide the cutoff for the sum over images. 
The causality-reversing property of the reflection mapping ensures that $G_{\Delta,0}$ is defined correctly on the entire boundary cylinder. Note, however, that because the isometry (\ref{nb-mov-iso}) shifts time as well as angle, in the general case of the moving particle, image points $b^{*n}$ or $b^{\#n}$ can be timelike-separated from $a$ even if points $a$ and $b$ are spacelike-separated. Also, if a point $x$ belongs to the living space, the point $x^R$ generally can be in the dead zone. Thus, the reflection geodesic is not generally a basic geodesic, but is a winding geodesic. In what follows we show that the winding of the reflection geodesic can be constructed using the same set of images that was used in construction of spacelike winding geodesics. 

\subsection{Windings of reflection geodesics}

Note that in case of static particle the identification (\ref{nb-mov-iso}) leads to just angular identification $\varphi \sim \varphi + \bar{\alpha}$. The full correlator is independent on the actual position of the wedge in the AdS$_3$ cylinder. Therefore, if $|\varphi_{a^R}-\varphi_{b}| < \bar{\alpha},\, \text{mod}(2\pi),$ then we can rotate the wedge to embed the entire reflection geodesic in the living space. If the inequality  is not satisfied, one has to use the winding geodesic. The same goes for the case of the moving defect for any values of angles. 

\begin{figure}
   \centering
\includegraphics[width=10cm]{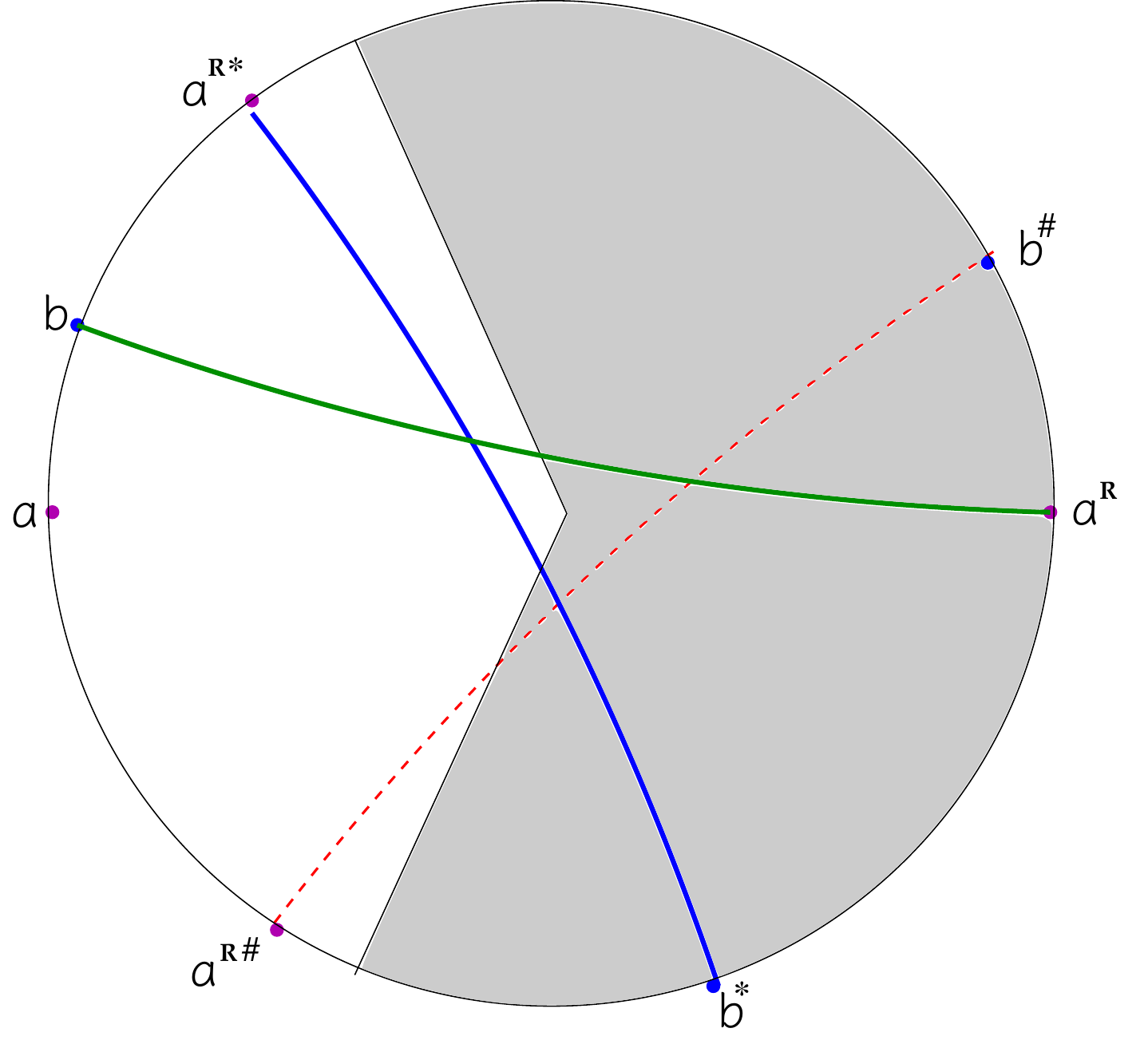}
\caption{The schematic plot of winding geodesic connecting the points $a^{R*}$ and $b$. The length of geodesic between $a^{R\#}$ and $b$ doesn't equal to the length of original reflection geodesic between $a^R$ and $b$. Here $\bar{\alpha}=5\pi/6$, static particle case.
} \label{2pi3winding}\end{figure}

The winding for the reflection geodesic is constructed as follows. 
Assume the position of the wedge is fixed. The basic geodesic has two parts: one of them lies in the living space and the other is in the dead zone. We can do the isometry transformation ($*$ or $\#$) $n$ times for both points and eventually the point $a^{R n*}$ or $a^{R n\#}$ will be in the living space. For instance if the angle of living space equals to $5\pi/6$ then $n=1$ (see Fig.\ref{2pi3winding}). Only one of the geodesics gives the right answer for the length of the basic geodesic. In the Fig.\ref{2pi3winding} it is the geodesic between $a^{R*}$ and $b^*$.

It is clear to generalize the previous case to arbitrary angle $\alpha$. If $n$ doesn't equal to one we get a few windings are build as follows. Full geodesic consists of many parts of supporting geodesics that connect $a^{R}$ and $b$, $a^{R *}$ and $b^*$, \dots, $a^{R n*}$ and $b^{n*}$ or  $a^{R}$ and $b$, $a^{R \#}$ and $b^{\#}$, ..., $a^{R n\#}$ and $b^{n\#}$. We also note that since we do not introduce new exits to the boundary, there are no new renormalization factors needed.

To take into account contributions of reflection geodesic from $(*, \#)$-images, we act in the similar manner. First, we evaluate the $\Theta$-functions and count all the images that actually contribute. Next, we check for every image point its separation from the point $a$, whether it is timelike or spacelike. Spacelike image geodesics are accounted for as they are. For those images, which are timelike-separated from $a$, we act via the reflection mapping and construct  winding for the reflection geodesic in the way described above for every image. 

\subsection{Examples}

\subsubsection{Static particle}

First, consider an important special case. In case of static defect when the deficit angle is $2\pi /N$ with $N\in \mathbb{Z}$, the spacetime is an orbifold AdS$_3/\mathbb{Z}_N$, the orbit of the isometry is a finite set of $N$ images. Therefore, two sums in (\ref{g-main}) merge into one finite sum which cycles through all $N$ images and all surviving $\Theta$-functions are equal to $1$ everywhere in the living space:
\be
\left< \mathcal{O}_\Delta(t, \vartheta) \mathcal{O}_\Delta(0, 0) \right> = \sum_{k=0}^{N-1}\, \left(\frac{1}{2\left(\cos t- \cos \left(\vartheta+2\pi \frac{k}{N}\right)\right)}\right)^{\Delta}\,. \label{geodesicsorbifold}
\ee
This result corresponds to the conformal field theory on a $\mathbb{Z}_N$-orbifold, as well as to the expression obtained from the traditional GKPW prescription \cite{AK}. 

\begin{figure}[t]
\centering
     \includegraphics[width=7cm]{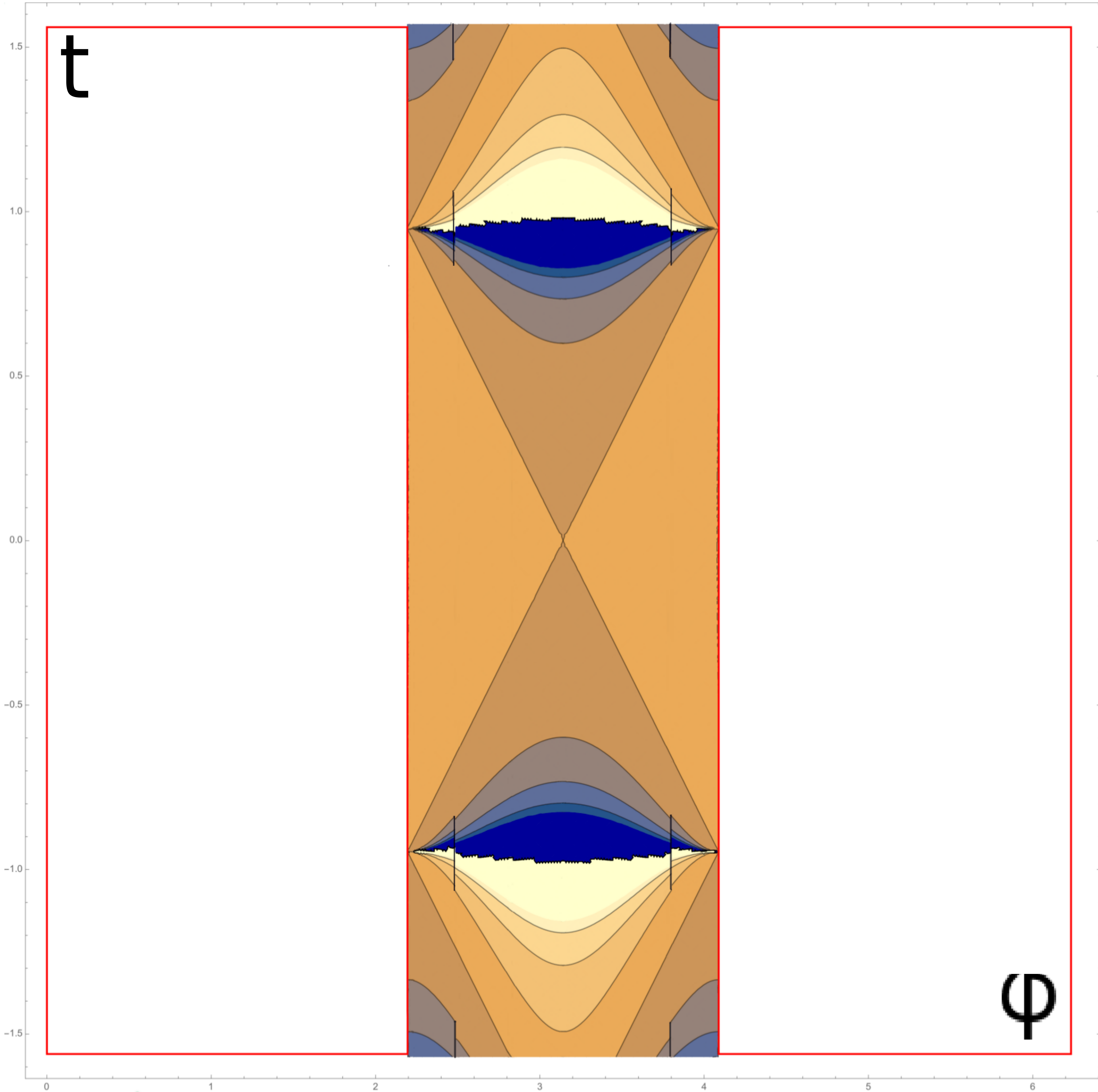}A.
     \includegraphics[width=7cm]{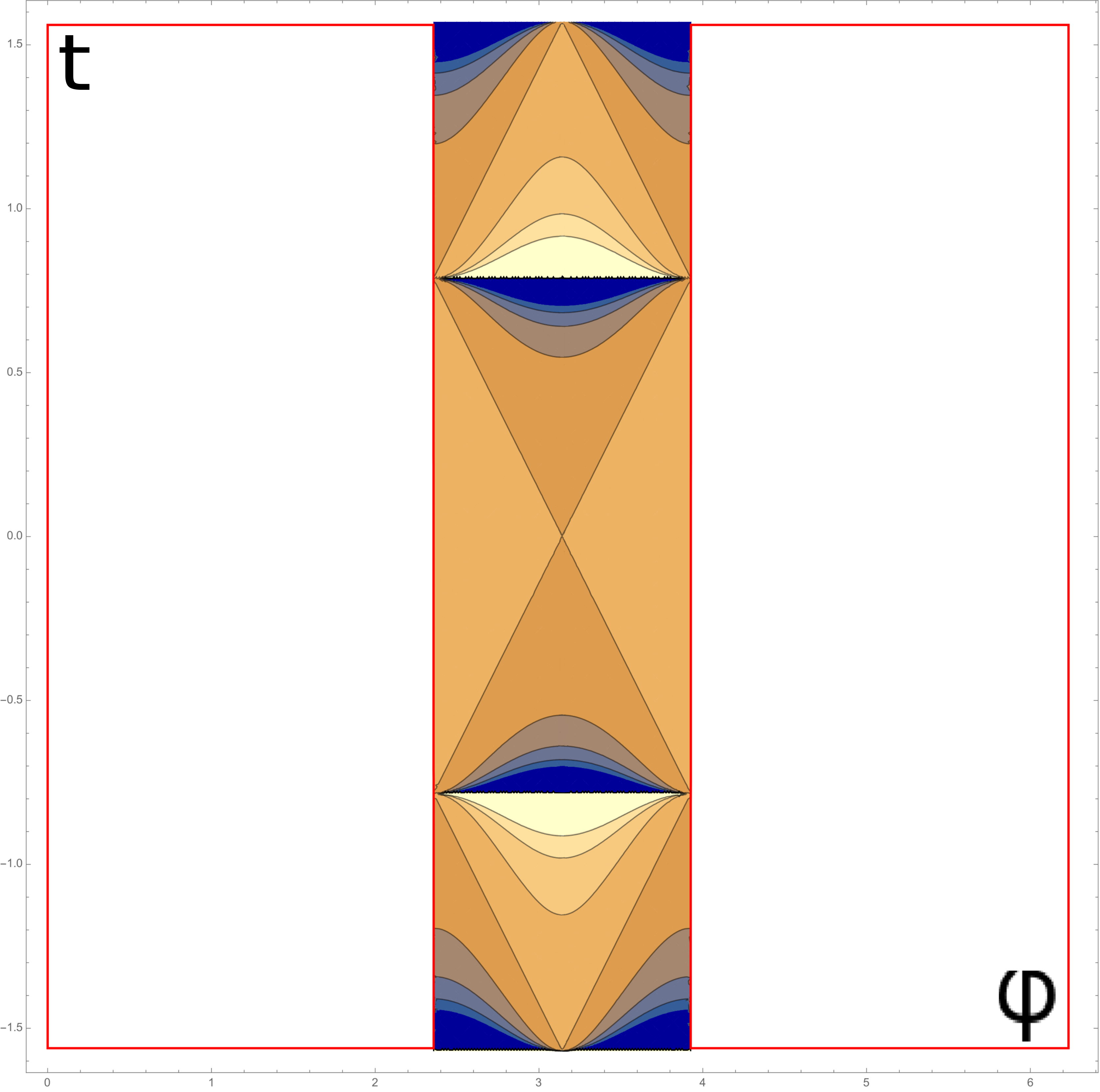}B.
       \caption{\textbf{A}. Density plot of inverse correlation function $G^{-1}(\varphi,t)$. Parameter values are $\varphi_a=\pi$, $t_a=0$; $\alpha=4\pi/3+0.2$, static case.\\\textbf{B}. Density plot of inverse correlation function $G^{-1}(\varphi,t)$. Parameter values are $\varphi_a=\pi$, $t_a=0$; $\alpha=3\pi/2$, static case.}\label{Static-dens}
\end{figure}

\begin{figure}
\centering
     \includegraphics[width=7cm]{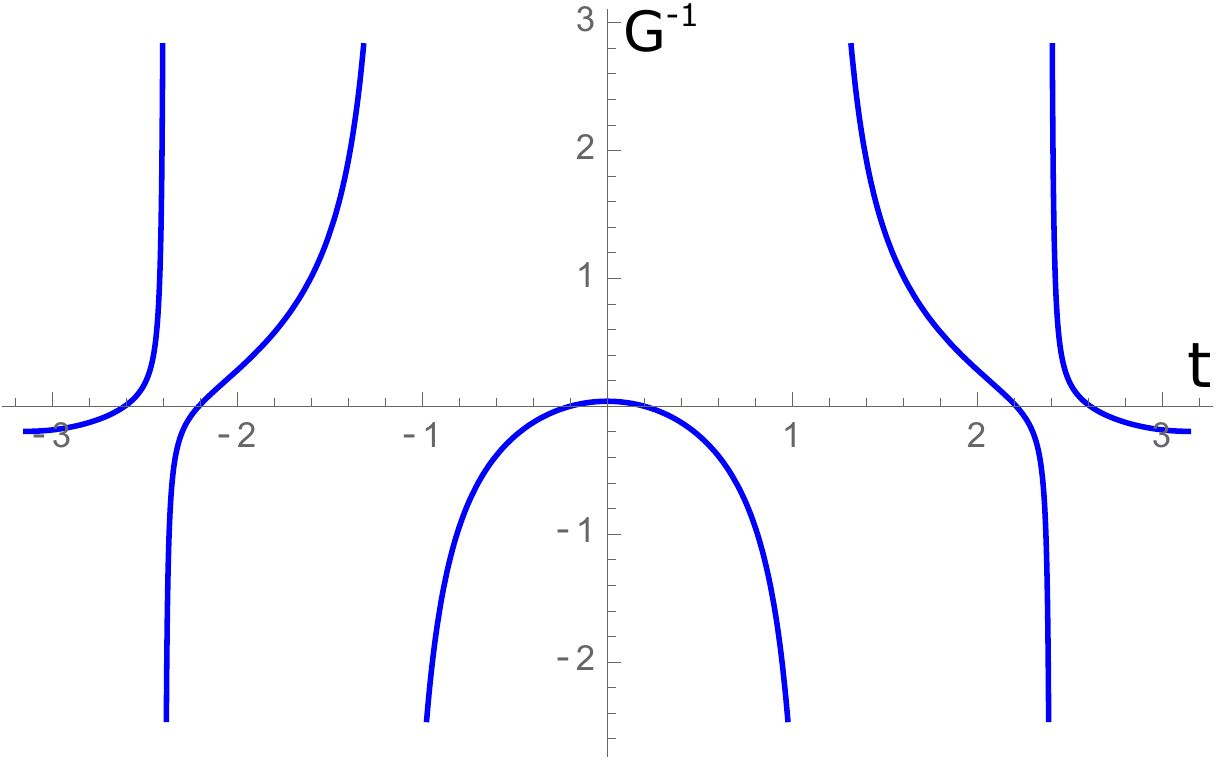}A. 
     \includegraphics[width=7cm]{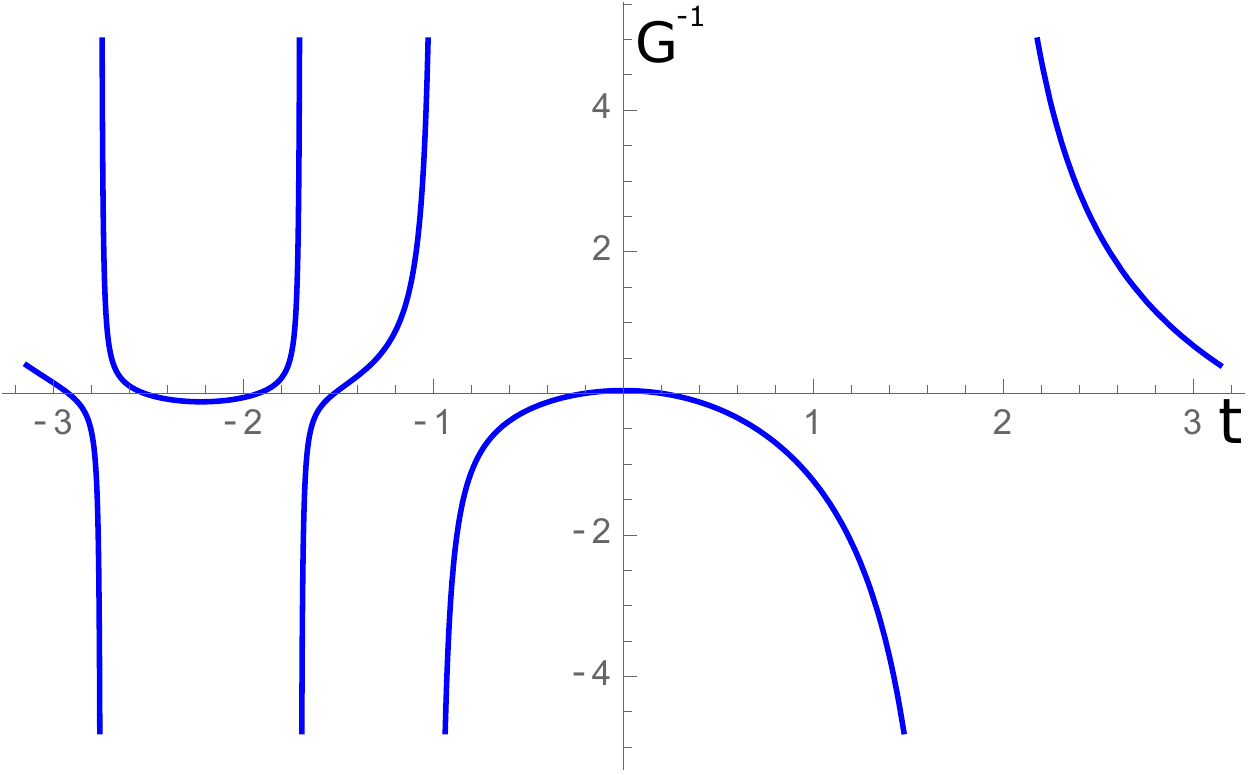}B.
       \caption{\textbf{A}. Plot of inverse correlation function $G^{-1}(t)$ in the static case. Parameter values are $\varphi_a=\pi$, $t_a=0$; $\varphi_b=\pi+0.2$; $\bar{\alpha}=2.3\pi/3$.\\\textbf{B}. Plot of inverse correlation function $G^{-1}(t)$. Parameter values are $\varphi_a=\pi$, $t_a=0$; $\varphi_b=\pi+0.2$; $\bar{\alpha}=2.3\pi/3$, $\xi=0.6$.}\label{2d}
\end{figure}

In the general case of static particle it is possible to perform the analytic continuation to the Euclidean background, and one can obtain the geodesics prescription in the full Lorentzian space by making the reverse rotation (of course, in this case one needs to specify the Lorentzian Green's function by introducing the $i\epsilon$ insertion by hand). Our prescription is consistent with this continuation. One finds that in non-orbifold case $\Theta$-functions generate discontinuities, which result into zone structure of geodesic correlators \cite{AAT}. This is an apparent artefact of our image method, which happens because we do not take into account corrections from the gravitational interaction with the particle \cite{AK}. The comparison of orbifold and non-orbifold cases is shown in Fig.\ref{Static-dens}. The time dependence on the interval $t \in [-\pi,\ \pi]$ is shown in Fig.\ref{2d}A.

\begin{figure}[t]
\centering
\includegraphics[width=7cm]{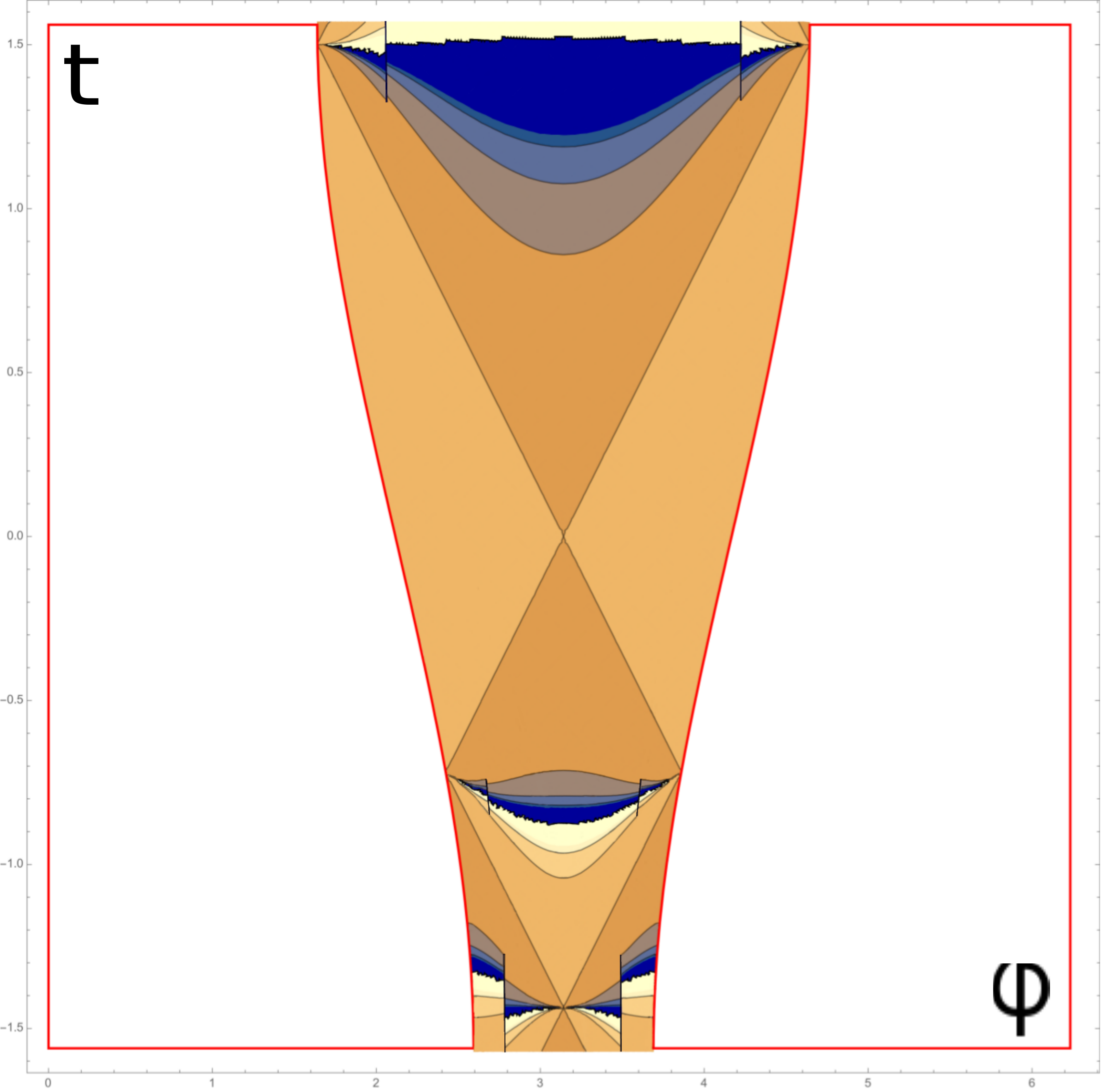}A.
\includegraphics[width=7cm]{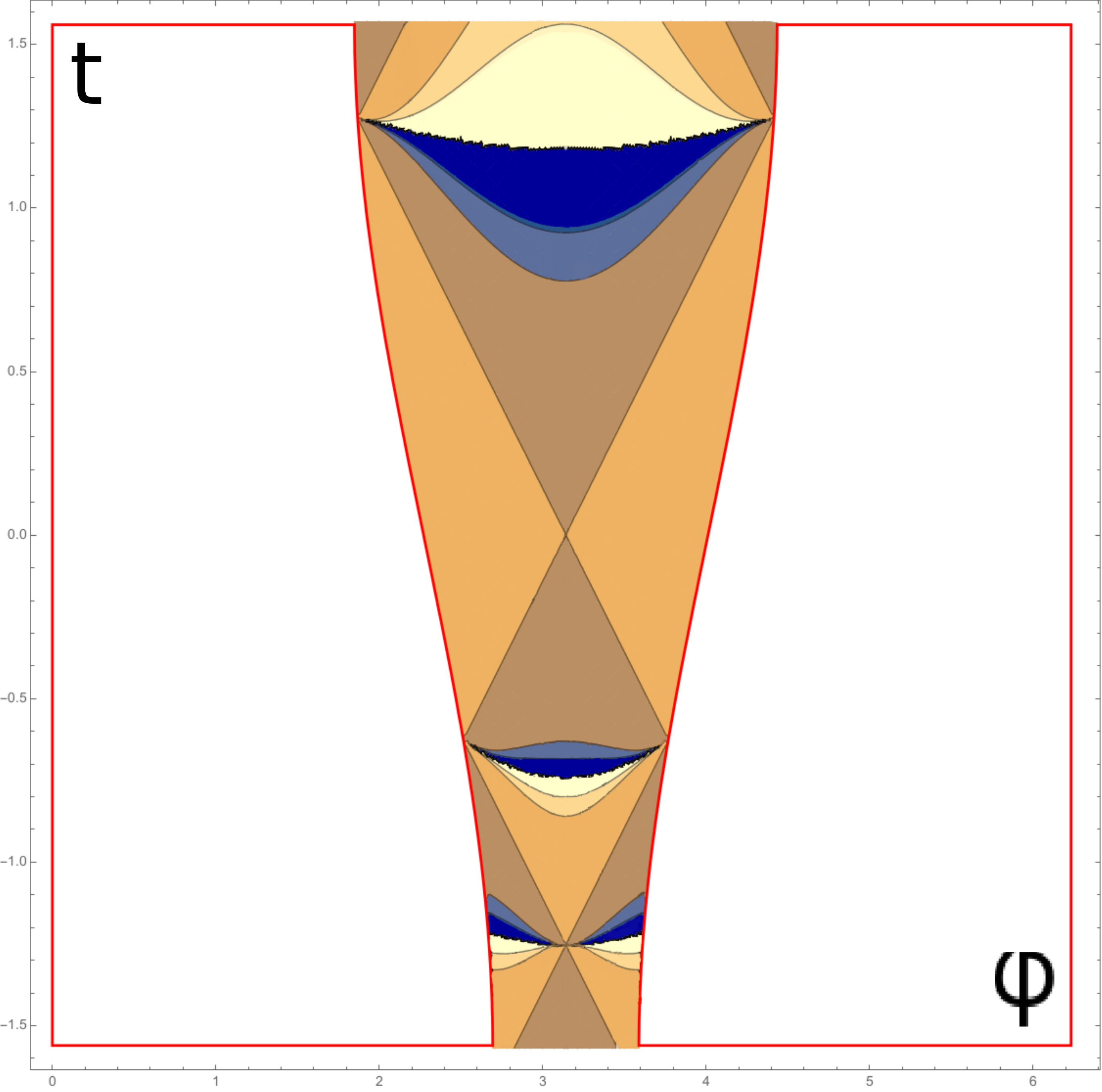}B.       
       \caption{\textbf{A}. Density plot of inverse correlation function $G^{-1}(\varphi, t)$. Parameter values are $\varphi_a=\pi$, $t_a=0$; $\alpha=4\pi/3+0.2$ and $\xi=0.6$.\\\textbf{B}.Density plot of inverse correlation function $G^{-1}(\varphi, t)$. Parameter values are $\varphi_a=\pi$, $t_a=0$; $\alpha=3\pi/2$ and $\xi=0.6$.}\label{Moving-dens}
\end{figure}

\subsubsection{Moving particle}

For the moving particle case, we observe (see Fig.\ref{Moving-dens})  that it is the central zone that shrinks when the living space decreases. Singularities also become more dense when the living space is small, which is consistent with the consideration of the static case with different sizes of the wedge. The time dependence also reflects this, see Fig.\ref{2d}B.

\section{Generalizations of the image method \label{Generals}}
\subsection{Multiple particles in the bulk}

Our images prescription for the correlator can be generalized to the case when the bulk spacetime is deformed by several particles\footnote{In \cite{AA} the particular case of two colliding ultrarelativistic particles was considered (in terms of spacelike correlations). }. Suppose that there are $P$ particles, inducing identification isometries $*_1$, \dots, $*_P$ (together with corresponding $\#$'s). Then the image method for correlators will schematically read (where $A=W,c$, or $ret$)
\bea\label{g-main-P}
&& G_\Delta^A (a, b)  = G^A_{\Delta,0} (a ,b)\,\Theta_{0}(a,b)\\&&+\sum_{\sigma\in S^P} \sum_{n_1, \dots, n_P}\sum_{\star = *,\#} \,G^A_{\Delta,n}(a,b^{\star^{n_1}_{\sigma(1)}\times\dots\times \star^{n_P}_{\sigma(P)}})\,\boldsymbol{\Theta}(a,b^{\star^{n_1}_{\sigma(1)}\times\dots\times \star^{n_P}_{\sigma(P)}})\,;
\eea
In this case we have to sum over orbits of all combinations of all $P$ isometries and their inverse. The $\boldsymbol{\Theta}$-function is a product of $\Theta_{\pm}^{(k)}$-functions (with $k$ enumerating the $\star$-isometries appearing in the $G$'s), which are just $1$-particle theta-functions defined for $k$-the wedge in the same way as in \ref{Geodesics}. The index sign of $\Theta_{\pm}^{(k)}$ is plus if the isometry index $\star_k = *_k$, and minus if  $\star_k = \#_k$. 

\subsection{Holographic entanglement entropy and images}

Here we explain how our images prescription can be used for the calculation of the holographic entanglement entropy (HEE). In particular, it useful to trace the temporal behaviour of HEE in non-stationary backgrounds, such as the background of moving massive particle. Note that for the case of moving massless particle the HEE was calculated in \cite{AA}.

The formula for HEE of the boundary region between points $a$ and $b$ in our case is 
\be
S(a, b) = \min \left\{ \begin{array}{ccc}
\mathcal{L}_{\text{ren}}(a, b)\,;\nn\\
\mathcal{L}_{\text{ren}}(a, b_K^*)\,;\\
\mathcal{L}_{\text{ren}}(a, b_L^\#)\,.\\
\end{array}\right.
\ee
Here $K$ and $L$ are minimal integers, such that $\Theta_{+}(a;\ b^*_{K})=1$ and $\Theta_{-}(a;\ b^\#_{L})=1$.
The entropy function has to choose the shortest renormalized geodesic of the basic geodesic and two winding geodesics. The image geodesics which represent windings are not equal-time geodesics in case of moving particle. For some values of parameters $\xi$ and $\alpha$, the interval between $a$ and $b_K^*$ (or $b_L^\#$) can become timelike, if the spatial separation is sufficiently small. However, despite apparent similarities, the situation here is slightly different than in case of two-point correlators. One has to keep in mind that image geodesics are just convenient objects to calculate the lengths of winding geodesics. The winding geodesics which compete in the calculation of the entropy are always equal-time geodesics, and they should be accounted for with no regard to the movement of the particle because of Lorentz-covariance in the bulk. The absence of apparent contribution from the timelike-separated points in the case of HEE is purely an artefact of the images prescription, whereas  in case of correlators it is also the artefact of the geodesic approximation itself. 

With that in mind, it is now clear that one has to find a way to keep the contribution from windings in the minimizing competition for entropy if some image points are pushed into the timelike region. Recall that the geodesic lengths between points $x$ and $y$ in AdS$_3$ is:
\be
\mathcal{L} = \ln (2(\cos (t_x - t_y) - \cos (\varphi_x - \varphi_y )))\,.
\ee
If the interval between $x$ and $y$ is spacelike, the difference between two cosines is positive, and if the interval is timelike, the difference is negative, and thus denies a real-valued answer for the geodesic length expression. Now, if $x$ and $y$ are timelike-separated, then $x$ and $y^R$ are spacelike separated, with $y^R$ being the reflection of $y$ as defined in (\ref{reflection}). Therefore, we adjust our images prescription so that if points $x$, $y$ are timelike-separated, than the corresponding contribution to the entropy is computed as (renormalized) geodesic length $\mathcal{L}_{\text{ren}}(x, y^R)$. In the form more suitable for calculations, the HEE formula than reads (with $t_a=t_b = t$):
\be
S(a, b) = \min \left\{ \begin{array}{ccc}
\ln (2(1 - \cos (\varphi_a - \varphi_b )))\,;\label{HEE}\\
\ln (2|\cos (t - t^*) - \cos (\varphi_a - \varphi_{b, K}^* )|)\,;\\
\ln (2|\cos (t - t^\#) - \cos (\varphi_a - \varphi_{b, L}^\# )|)\,.\\
\end{array}\right.
\ee
This formula allows to analyze the dependence of HEE on the background of moving particle(s) of time $t$ continuously for any length of time\footnote{A similar result for the HEE was obtained from the CFT side in \cite{Asplund2014} in the context of a specific local quench at the boundary}.

\section{Conclusion \label{conclusion}}

We have proposed the prescription to deal with breakdown issues of the geodesics images prescription in the Lorentzian AdS-deficit spacetime for calculation of the two-point boundary correlator and holographic entanglement entropy. The recipe allows to calculate these quantities in non-stationary defect backgrounds and is easily generalizable to other multi-connected locally AdS spacetimes.
\begin{itemize}
\item The geodesic two-point correlator is continued to the case where the points are timelike-separated by transitioning to reflection geodesics in the timelike region. The appropriate causal structure of Lorentzian correlators is re-established by hand in the form of additional factors of special form.
\item The prescription can be generalized to other physically interesting quotients of AdS$_3$, such as AdS$_3$ with multiple particles or BTZ black hole background and its deformations. It can be also used to calculate the holographic entanglement entropy and trace its time dependence in non-stationary backgrounds.
\end{itemize}
 
\appendix
\section*{Acknowledgements}

The authors are grateful to Dmitrii Ageev for useful discussions. This work was done in Steklov Mathematical Institute. I. A. and M. T. are supported by the Russian Science Foundation (project 14-11-00687).

\section{Renormalization of image geodesics in AdS with particles\label{renormalization}}

Here we briefly review the identification-invariant renormalization scheme used in  \cite{AAT} for geodesics on a cone that is required for the image method to be self-consistent. 
One renormalizes the lengths of geodesics reaching the boundary by subtracting the diverging part. This renormalization must be uniform for every geodesic in the sum, and it also must respect the isometry (\ref{nb-mov-iso}). This leads to the fact that lengths of the image geodesics recieve additional numerical renormalization factors compared to the length of the basic (the one that connects two points that are actual arguments of the correlator) geodesic. Thus, while the contribution of the latter to the correlator is defined as the usual AdS correlator $G_{\Delta,0}$, the images contributions are defined as 
\bea\label{G-ren-definition}
G_{\Delta,ren,n}(\varphi^*_{a,n},t^*_{a,n};\varphi_b,t_b)&\equiv &G_{\Delta,0}(\varphi^*_{a,n},t^*_{a,n};\varphi_b,t_b)Z_n(\varphi^*_{a,n},t^*_{a,n};\varphi_b,t_b)\,;\\\nn
G_{\Delta,ren,n}
( \varphi^\#_{a,n}, t^\#_{a,n};\varphi_b,t_b)&\equiv &G_{\Delta,0}
( \varphi^\#_{a,n}, t^\#_{a,n};\varphi_b,t_b)\, \bar{Z}_n( \varphi^\#_{a,n},t^{\#}_{a,n};\varphi_b,t_b)\,.\eea
The normalization factors $Z_n$ are combined from the renormalization coefficients which come action of the isometry: 
\bea\label{two-point-three-ren}
Z_n( t^{\#}_{a,n},\varphi^{\#}_{a,n};t_b,\varphi_b)&=&C_{a^{\#n}}^{-1/2}=C_{a^{\#(n-1)}}^{-1/2}C_{b^{*}}^{-1/2}=...=C_{b^{*n}}^{-1/2}\,;\\
\bar{Z}_n( t^{*}_{a,n},\varphi^{*}_{a,n};t_b,\varphi_b)&=&C_{a^{*}}^{-1/2}=C_{a^{*(n-1)}}^{-1/2}C_{b^{\#}}^{-1/2}=...=C_{b^{\#n}}^{-1/2}\,;
\eea
where $C$-coefficients are defined from
    \bea\nn
C_{b^{*n}}&=&\left(\mathcal{B}_{\xi}(n\alpha) \cos \varphi_b+\sin t_b (1+2\sinh^2\xi\sin^2\frac{n\alpha}{2})\right)^2+ \cos ^2t_b\,;\nn\\
C_{a^{\#n}}&=& \left(\mathcal{B}_{\xi}(-n\alpha) \cos \varphi_a+\sin t_a (1+2\sinh^2\xi\sin^2\frac{n\alpha}{2})\right)^2+ \cos ^2t_a\,.\eea
These renormalization factors turn into unit in the case of static particle $\xi = 0$. 
The image contributions to the correlator defined  by (\ref{G-ren-definition}) thus satisfy the isometry invariance property:
\bea\nn
G_{\Delta,ren,n}(\varphi^*_{a,n},t^*_{a,n};\varphi_b,t_b)&=&G_{\Delta,ren,n}(\varphi^*_{a,n-1},t^*_{a,n-1}; \varphi^\#_{b,1}, t^\#_{b,1})=G_{\Delta,ren,n}(\varphi_a,t_a; \varphi^\#_{b,n}, t^\#_{b,n})\,;\\\nn
G_{\Delta,ren,n}(\varphi^\#_{a,n},t^\#_{a,n};\varphi_b,t_b)&=&G_{\Delta,ren,n}(\varphi^\#_{a,n-1},t^\#_{a,n-1}; \varphi^{*}_{b,1}, t^{*}_{b,1})=G_{\Delta,ren,n}(\varphi_a,t_a; \varphi^*_{b,n}, t^*_{b,n})\,.
\eea

\end{document}